\documentclass[a4paper]{jpconf}
\usepackage{graphicx}
\begin{document}
\title{Future of superheavy element research: Which nuclei could be synthesized within the next few years?}

\author{Valeriy Zagrebaev}
\address{Flerov Laboratory of Nuclear Reactions, JINR, Dubna, Moscow Region, Russia}
\ead{zagrebaev@jinr.ru}
\author{Alexander Karpov}
\address{Flerov Laboratory of Nuclear Reactions, JINR, Dubna, Moscow Region, Russia}
\ead{karpov@jinr.ru}
\author{Walter~Greiner}
\address{Frankfurt Institute for Advanced Studies, J.W. Goethe-Universit\"{a}t, Frankfurt, Germany}
\ead{greiner@fias.uni-frankfurt.de}

\begin{abstract}
Low values of the fusion cross sections and very short half-lives of nuclei with Z$>$120 put obstacles in synthesis of new elements.
Different nuclear reactions (fusion of stable and radioactive nuclei, multi-nucleon transfers and neutron capture),
which could be used for the production of new isotopes of superheavy (SH) elements, are discussed in the paper.
The gap of unknown SH nuclei, located between the isotopes which were produced earlier in the cold and hot fusion reactions,
can be filled in fusion reactions of $^{48}$Ca with available lighter isotopes of Pu, Am, and Cm.
Cross sections for the production of these nuclei are predicted to be rather large,
and the corresponding experiments can be easily performed at existing facilities.
For the first time, a narrow pathway is found to the middle of the island of stability owing to possible
$\beta^+$-decay of SH isotopes which can be formed in ordinary fusion reactions of stable nuclei.
Multi-nucleon transfer processes at near barrier collisions of heavy (and very heavy, U-like) ions
are shown to be quite realistic reaction mechanism allowing us to produce new neutron enriched heavy nuclei
located in the unexplored upper part of the nuclear map.
Neutron capture reactions can be also used for the production of the long-living neutron rich SH nuclei.
Strong neutron fluxes might be provided by pulsed nuclear reactors and by nuclear explosions in laboratory conditions
and by supernova explosions in nature. All these possibilities are discussed in the paper.
\end{abstract}

\section{Motivation}

\begin{figure}[h]
\begin{center}
\includegraphics[width = 15.0 cm, bb=0 0 2113 682]{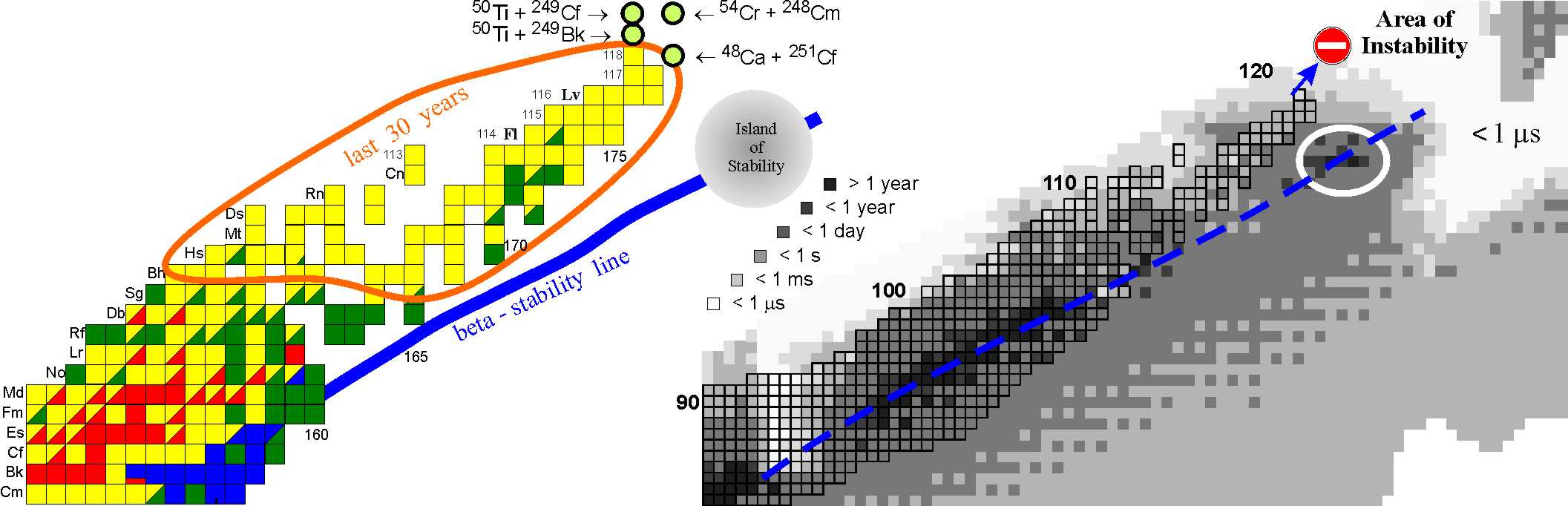}
\end{center}
\caption{\label{mapup1}(Left panel) Upper part of the nuclear map. Current and planned experiments
on synthesis of elements 118-120 are shown. (Right panel) Predicted half-lives of SH nuclei and the ``area of instability''.
Known nuclei are shown by the outlined rectangles.}
\end{figure}

Significant progress has been achieved during the last thirty years in synthesis of superheavy nuclei
using the ``cold'' \cite{Hofmann00,Morita07} and ``hot'' ($^{48}$Ca induced) \cite{Ogan07} fusion reactions (see Fig.\ \ref{mapup1}).
The heaviest element 118 was synthesized with the cross section of about 1~pb in fusion of $^{48}$Ca with
heaviest available target of $^{249}$Cf \cite{118}.
A kind of ``world record'' of 0.03~pb in production cross section of element 113 has been obtained
in this field within more than half-year irradiation of $^{209}$Bi target with $^{70}$Zn beam \cite{Morita07}.

Due to the bending of the stability line toward the neutron axis, in fusion reactions of stable nuclei one may produce
only proton rich isotopes of heavy elements. That is the main reason for the impossibility to reach the center of
the ``island of stability'' ($Z\sim 110\div 120$ and $N\sim 184$) in fusion reactions with stable projectiles.
Note that for elements with $Z > 100$ only neutron deficient isotopes (located to the left of the stability line) have been synthesized so far
(see the left panel of Fig.\ \ref{mapup1}).

Further progress in the synthesis of new elements with $Z > 118$ is not quite evident.
Cross sections of the ``cold'' fusion reactions decrease very fast with increasing charge of the projectile.
(they become less than 1~pb already for $Z\ge 112$ \cite{Hofmann00,Morita07}).
For the more asymmetric $^{48}$Ca induced fusion reactions rather constant values (of a few picobarns) of the cross sections
for the production of SH elements with $Z=112\div 118$ have been predicted in \cite{Z03,Z04}.
This unusual (at first sight) behavior of the cross sections was explained in \cite{Z03,Z04} by the relatively slow decrease
of the fusion probability (in contrast to the more symmetric  ``cold'' fusion reactions) and by the increasing survival probability
of compound nuclei (CN) owing to increasing values of their fission barriers caused by the larger shell corrections as the CN approach the neutron
and proton closed shells \cite{Moller97,Sob03} in the region of the island of stability.
These predictions have been fully confirmed by the experiments performed in Dubna \cite{Ogan07} and later in Berkeley \cite{114_5n}
and at GSI \cite{GSI_114,GSI_116}.

For the moment $^{249}_{98}$Cf (T$_{1/2}$ = 351 yr) is the heaviest available target that can be used in experiments.
The half-life of the einsteinium isotope, $^{254}_{99}$Es, is 276 days, sufficient to be used as target material.
In principle, this isotope might be produced in nuclear reactors, but it is rather difficult to accumulate the required amount
of this matter (several milligrams) to prepare a target.
The cross section for the production of element 119 in the hypothetical $^{48}$Ca+$^{254}$Es fusion reaction
is about 0.3~pb \cite{Zag12} which is more promising as compared with the $^{50}$Ti+$^{249}$Bk fusion reaction \cite{Zag08}.
Anyhow, to get SH elements with $Z > 118$ in fusion reactions in a more realistic way, one should proceed to heavier than $^{48}$Ca projectiles.

Strong dependence of the calculated evaporation residue (EvR) cross sections for the production of SH elements on the mass asymmetry
in the entrance channel makes the nearest to $^{48}$Ca projectile, $^{50}$Ti, most promising for further
synthesis of SH nuclei. Our calculations demonstrated that the use of the titanium beam instead of $^{48}$Ca
decreases the yield of the same SH element due to a worse fusion probability by about factor 20 \cite{Zag08}.
The elements 119 and 120 can be produced in the fusion reactions of $^{50}$Ti with $^{249}$Bk and $^{249}$Cf targets
(or in the $^{54}$Cr+$^{248}$Cm fusion reaction) with the cross sections of about 0.04~pb \cite{Zag08} which are
already at the limit of the experimental possibilities. The synthesis of these nuclei may encounter also another important problem.
The proton rich isotopes of SH elements produced in these reactions are rather short-living due to large values of $Q_\alpha$.
Their half-lives are very close to the critical value of one microsecond needed for the CN to pass through the separator up to the focal plane detector.
The next elements (with $Z > 120$) being synthesized in such a way might be already beyond this natural time limit for their detection
(see the right panel of Fig.\ \ref{mapup1}).

Thus, future studies of SH elements are obviously connected with the production of neutron enriched and longer living isotopes of SH nuclei.
The possibilities of using fusion reactions (including radioactive beams), multi-nucleon transfer reactions and neutron capture processes for this purpose
are discussed in the paper (see also \cite{Zag12,Zag08,Zag11,Zag11b}).

\section{Fusion, multi-nucleon transfer and neutron capture processes}

There are only three methods for the production of transuranium elements, namely, fusion of heavy nuclei, multi-nucleon transfer reactions
and a sequence of neutron capture and $\beta^-$ decay processes. The cross section of SH element production in a heavy ion fusion reaction
(with subsequent evaporation of {\it x} neutrons in the cooling process) is calculated as follows:

\begin{equation}
\sigma _{\rm EvR}^{\rm xn}(E)=\frac{\pi}{k^2} \sum\limits_{l =
0}^\infty (2l + 1)P_{\rm cont}(E,l)\cdot P_{\rm CN}(E^*,l) \cdot
P_{\rm xn} (E^*,l). \label{Sig}
\end{equation}

The empirical channel coupling model \cite{Zag02} can be successfully used to calculate the penetrability of the
multi-dimensional Coulomb barrier $P_{\rm cont}(E,l)$ and the corresponding capture cross section, $\sigma _{\rm
cap}(E)=\pi/k^2 \sum (2l+1) P_{\rm cont}$. The standard statistical model is usually used for the calculation of the survival
probability $P_{\rm xn}(E^*)$ of an excited CN. All the decay widths may be easily calculated directly at the web site \cite{NRV}
by the Statistical Model Code of NRV.

The calculation of the probability for the CN formation in competition with the quasi-fission process, $P_{\rm CN}(E^*,l)$, is the most difficult problem.
In a well studied case of near-barrier fusion of light and medium nuclei, when a fissility of CN is not so high,
the fusing nuclei overcoming the potential barrier form a compound nucleus with a probability close to unity,
i.e., $P_{\rm CN}=1$, and, thus, this reaction stage does not influence the yield of EvR at all.
In the fusion of very heavy ions, the system of two touching nuclei may evolve with a high probability directly into the exit fission channels
without CN formation, which means that the so-called process of ``fast fission'' or quasi-fission takes place \cite{Toke85}.
At incident energies around the Coulomb barrier in the entrance channel the fusion probability $P_{\rm CN}\sim 10^{-3}$
for mass asymmetric reactions induced by $^{48}$Ca and much less for more symmetric combinations used in the ``cold'' synthesis \cite{Z03}.

\begin{figure}[h]
\begin{center}
\includegraphics[width = 12.0 cm, bb=0 0 2680 1521]{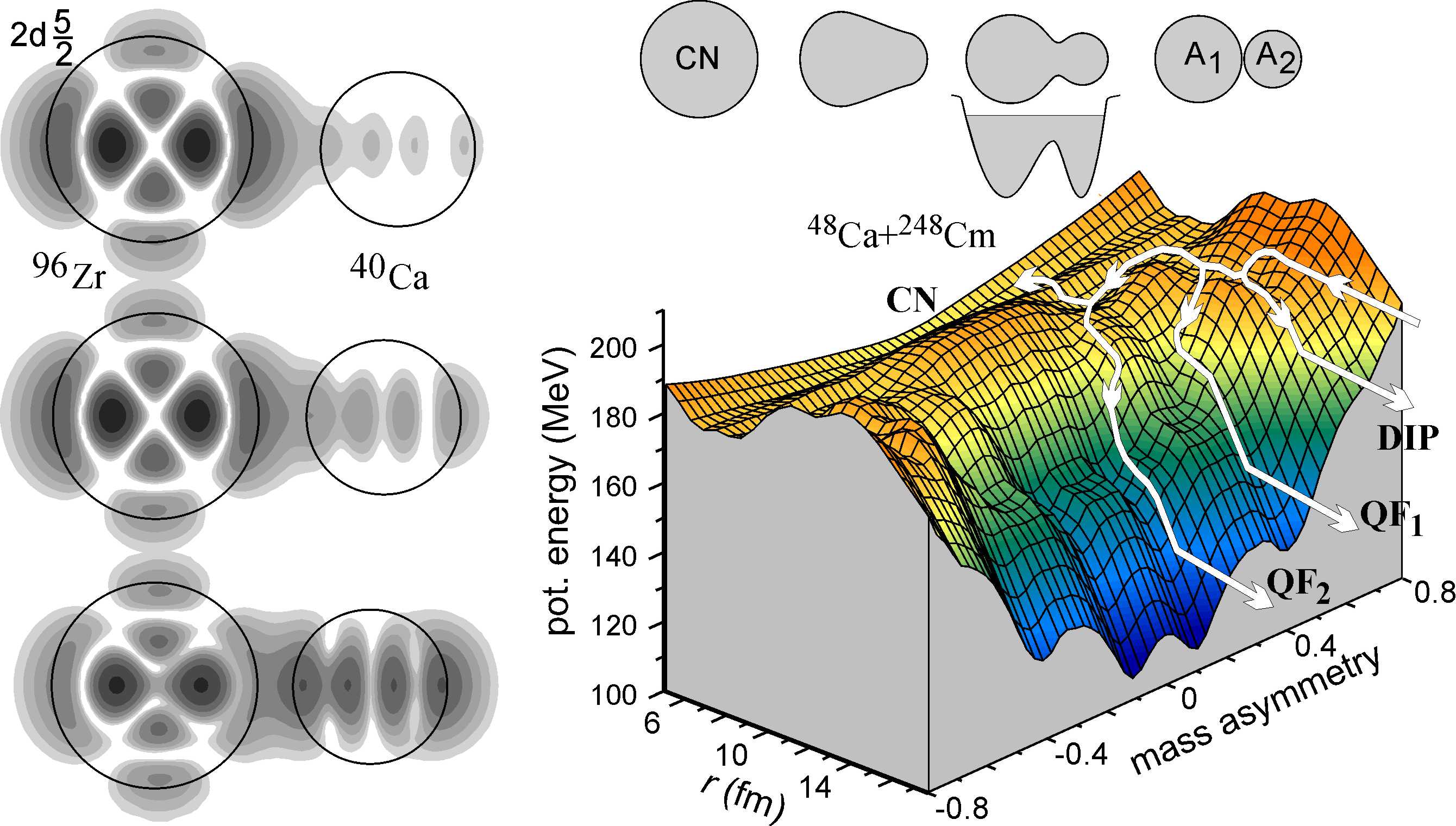}
\end{center}
\caption{\label{adiabat}(Left panel) Amplitude of the wave function of valence neutron initially located in the 2d state of $^{96}$Zr
nucleus approaching a $^{40}$Ca nucleus in a head-on collision at near-barrier center-of-mass energy E = 97~MeV.
(Right panel) Adiabatic potential energy of $^{48}$Ca+$^{248}$Cm nuclear system calculated within the two-center shell model.
Different reaction channels (deep inelastic scattering, quasi-fission and fusion) are shown schematically by the arrows.}
\end{figure}

At near barrier collisions, relative motion of heavy ions is very slow. In this case, much faster moving nucleons of colliding nuclei
have enough time to adjust their motion over the volumes of two nuclei forming a two-center mono-nucleus, i.e., the wave functions of valence nucleons
follow the two-center molecular states spreading over both nuclei. Such behavior of nucleons is confirmed by explicit solution
of the time-dependent Schr\"{o}edinger equation \cite{ZSG07}. An example is shown in the left panel of Fig.\ \ref{adiabat} for the case
of near barrier collision of $^{40}$Ca with $^{96}$Zr. The same fusion dynamics with a neck formation was found also within TDHF calculations \cite{Umar08,Simenel12}.

Thus, the two-center shell model looks quite appropriate to describe adiabatic evolution of heavy two-center nuclear system transforming
from the configuration of two touching nuclei into the configuration of more or less spherical CN (fusion) or into the configuration of two deformed
re-separated fragments (dominating quasi-fission process). Elongation of the system (distance between nuclear centers) is the main collective degree of freedom
here, but the other variables (nucleon transfer, deformations, neck parameter) are also very important.
We used the extended version of the two-center shell model \cite{extTCSM} for the calculation of adiabatic multi-dimensional potential energy surface
(an example is shown in the right panel of Fig.\ \ref{adiabat} for the case of near barrier collision of $^{48}$Ca with $^{248}$Cm).

In collisions of heavy ions kinetic energy of relative motion transforms quickly into internal excitation (temperature) of nuclear system
and, thus, the fluctuations play a significant role. It was shown in Ref.~\cite{Zag05} that the multidimensional Langevin-type dynamical equations of motion
can be successfully used for simultaneous description of deep inelastic scattering (multi-nucleon transfer), quasi-fission and fusion processes
of low-energy nucleus-nucleus collisions. Simultaneous description of all these strongly coupled processes allows one to avoid big errors in absolute
normalization of the corresponding cross sections due to conservation of the total flux (no one collision event is lost).
All the cross sections are calculated in a quite natural way, just by counting the events coming into a given reaction channel.

\begin{figure}[ht]
\begin{center}
\includegraphics[width = 15.0 cm, bb=0 0 2331 704]{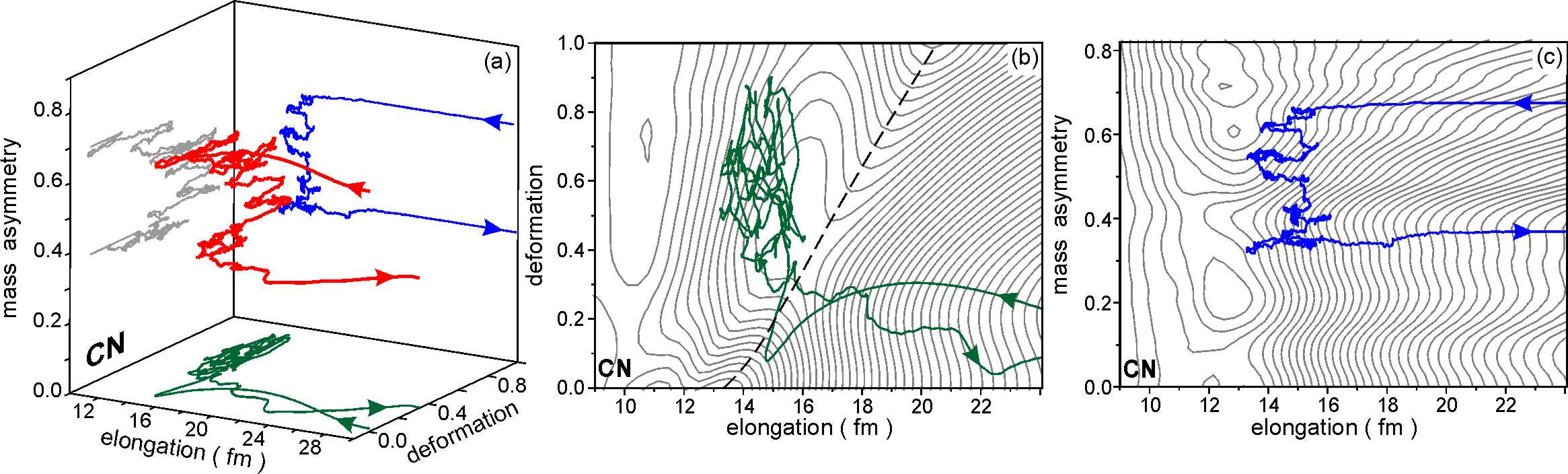} \end{center}
\caption{Collision of $^{48}$Ca +$ ^{248}$Cm at $E_{c.m.}=210$~MeV. A typical trajectory calculated within the Langevin equations and going to the
quasifission exit channel is shown in three-dimensional space (a) and projected onto the ``deformation--elongation'' (b)
and ``mass-asymmetry–-elongation'' (c) planes (along with the landscape of the potential energy).\label{traj}}
\end{figure}

In the case of the near-barrier collisions of heavy nuclei, only a few trajectories (of many thousands tested) reach the
CN configuration (small values of elongation and deformation parameters). All others go out to the dominating deep inelastic and/or
quasi-fission exit channels (see the right panel of Fig.\ \ref{adiabat}). One of such trajectories is shown in Fig.\ \ref{traj} in the three-dimensional space
of ``elongation-–deformation–-mass-asymmetry''. After nuclei approach each other their excitation energy (temperature) increases,
and fluctuations also increase.

The synthesis of heavier nuclei in the multiple neutron capture reactions with subsequent $\beta^-$ decay
is a well studied process (see, for example, \cite{Dorn62,Seaborg}).
Relative yields of the isotopes formed in this process may be found as a solution of the following set of differential equations
(somewhat simplified here):

\begin{eqnarray}\label{equation}
&\displaystyle {\frac{dN_{Z,A}}{dt}} = N_{Z,A-1} n_0 \sigma_{n\gamma}^{Z,A-1} - N_{Z,A} n_0 \sigma_{n\gamma}^{Z,A}
- N_{Z,A}[\lambda_{Z,A}^{\beta -} + \lambda_{Z,A}^{fis} + \lambda_{Z,A}^{\alpha}]\\ \nonumber &
+ N_{Z-1,A}\lambda_{Z-1,A}^{\beta -} + N_{Z+2,A+4}\lambda_{Z+2,A+4}^{\alpha},
\end{eqnarray}
where $n_0$ is the neutron flux (number of neutrons per square centimeter per second) and $\lambda_{Z,A}^i=ln2/T^i_{1/2}$ is the decay rate
of the nucleus $(Z,A)$ into the channel $i$ (i.e., $\beta^-$ and alpha decays and fission).
For simplicity, here we ignore the energy distribution of the neutrons and, thus, the energy dependence
of the neutron capture cross section $\sigma_{n\gamma}^{Z,A}$. Neutrons generated by fission
in nuclear reactors and in explosions are rather fast (far from the resonance region) and the neutron capture cross section
is a smooth function of energy with the value of about 1 barn.

The key quantity here is the time of neutron capture, $\tau_n = (n_0\sigma^{Z,A}_{n\gamma})^{-1}$.
If $\tau_n$ is shorter than the half-life of a given nucleus $T_{1/2}(Z,A)$ then the next nucleus $(Z,A+1)$ is formed by neutron capture.
Otherwise the nucleus $(Z,A)$ decays before it has time to capture next neutron.
In nuclear reactors typical value of $\tau_n\sim$1~year, and the nucleosynthesis occurs along the stability line by a sequence of neutron capture
and $\beta^-$ decay processes breaking at the short-living fissile fermium isotopes (see below).
In nuclear explosion $\tau_n\sim 1\, \mu$s, and more than 20 neutrons can be captured by a nucleus before it decays.

To solve Eq.(\ref{equation}) numerically one needs to know the decay properties of neutron rich nuclei which are not studied yet experimentally.
This significantly complicates any analysis of the multiple neutron capture processes. Details on our estimations of decay properties
of heavy and SH nuclei can be found in Ref.\ \cite{Karpov12}.

\section{Predictions and proposals for experiments}

\subsection{Elements 119 and 120 are coming}

\begin{figure}[ht]
\begin{center}
\includegraphics[width = 15.0 cm, bb=0 0 2161 640]{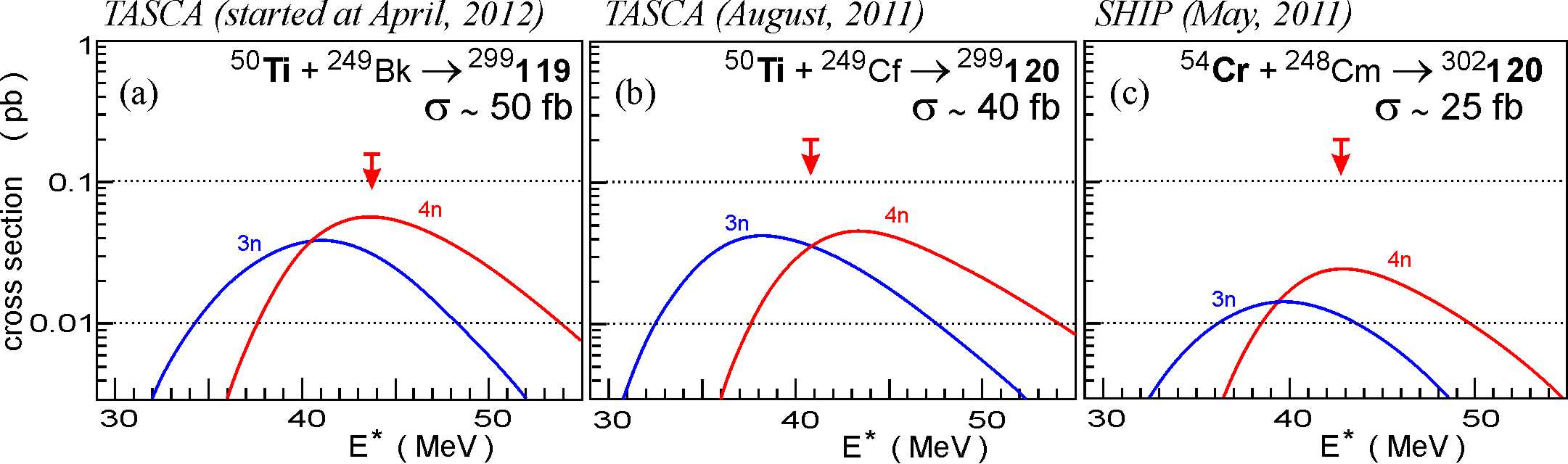} \end{center}
\caption{Predicted cross sections for synthesis of element 119 and 120 in the $^{50}$Ti + $^{249}$Bk (a),
$^{50}$Ti + $^{249}$Cf (b) and $^{54}$Cr + $^{248}$Cm (c) fusion reactions.
The arrows indicate the upper limits reached in the experiments performed at GSI by beginning of June 2012.
\label{120}}
\end{figure}

The fusion reactions of $^{50}$Ti with $^{249}$Bk and $^{249}$Cf targets (as well as $^{54}$Cr+$^{248}$Cm)
are the most promising for the production of new elements 119 and 120. The excitation functions for these reactions have been predicted several years ago \cite{Zag08}
(see Fig.\ \ref{120}). Two last reactions aimed on synthesis of element 120 were preliminary tested in 2011 at GSI by about one-month irradiations
of curium and californium targets with chromium \cite{SHIP120} and titanium \cite{TASCA120} beams, correspondingly. Experimental conditions
(beam intensity and short time of irradiation) allowed one to get only the upper limits of the cross sections for both reactions (see Fig.\ \ref{120}).
Experiment on the production of element 119 in the $^{50}$Ti+$^{249}$Bk fusion reaction was started in May of 2012 at GSI with the use
of separator TASCA. According to the predicted cross section, the first event could be detected within about 5 months of irradiation
(if the existing intensity of Ti beam will be maintained). Approximately the same time of irradiation is needed to produce element 120.

Production of elements with Z$>$120 is not so evident. There are two main objections. First, the EvR cross sections for the production of SH elements
become extremely low for the fusion reactions of projectiles heavier than $^{48}$Ca with available actinide targets. Second, half-lives of the isotopes
of SH elements with Z$>$120 produced in these reactions could be shorter than a few microseconds, time needed for SH nucleus to pass through separator.
Thus, the production of SH elements with Z$>$120 looks rather vague in the nearest future. The use of radioactive ion beams does not solve this problem \cite{Zag08}.

\subsection{Filling the Gap of not-yet-synthesized isotopes of SH elements (Z=106-116)}

An important area of SH isotopes located between those produced in the cold and hot fusion reactions remains unstudied yet (see the gap in the upper part
of the nuclear map in Fig.\ \ref{mapup1}). Closeness of the island of stability (confirmed, for example, by the fact that the half-life of the
isotope $^{285}$Cn produced in the hot fusion reaction is longer by almost five orders of magnitude than the $^{277}$Cn isotope of the same element produced
in the cold synthesis) testifies about strong shell effects in this area of the nuclear map. Understanding these effects, as well as other properties of SH
nuclei, is impeded significantly by the absence of experimental data on decay properties of the not-yet-synthesized isotopes
of already-known SH elements. Knowledge of the trends (especially along the neutron axis) of all decay properties of
these nuclei (fission, $\alpha$- and $\beta$-decays) may help us to predict more accurately the properties of SH nuclei located at (and to
the right of) the line of beta-stability, including those that are located in the island of stability.

\begin{figure}[ht]
\begin{center}
\includegraphics[width = 12.0 cm, bb=0 0 2306 737]{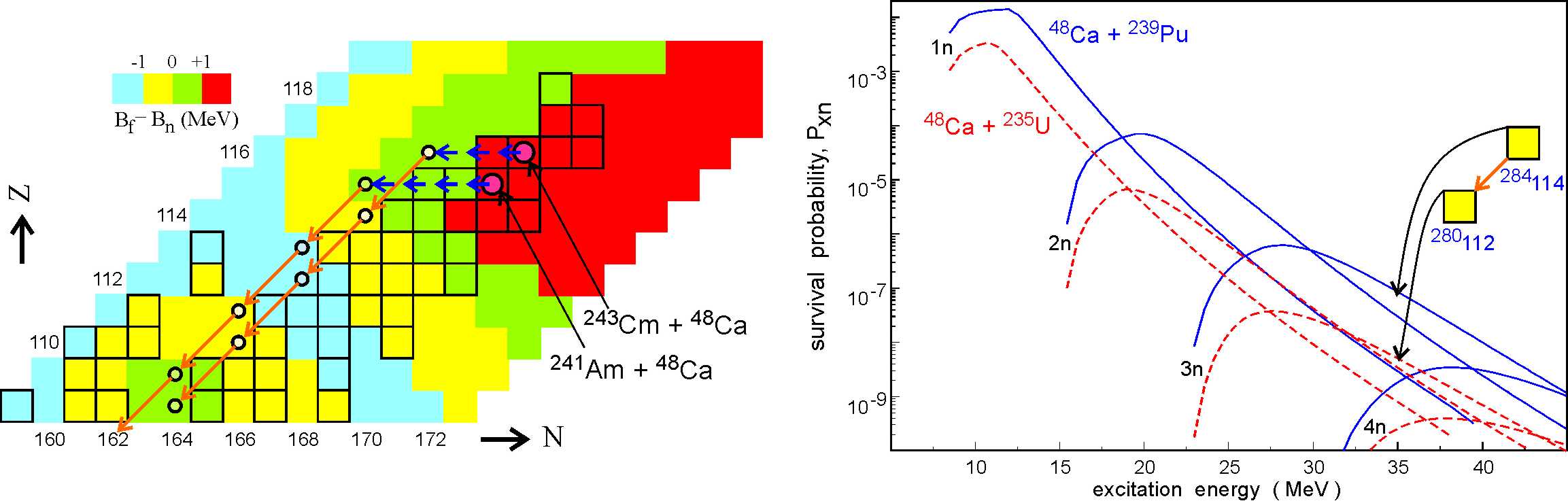} \end{center}
\caption{Left panel: The values of $B_f-B_n$ as a function of proton and neutron numbers.
Known isotopes of SH elements are marked by the bordered rectangles.
As an example the compound nuclei $^{289}$115 and $^{291}$116, formed in fusion reactions $^{48}$Ca+$^{241}$Am and $^{48}$Ca+$^{243}$Cm,
are shown along with $\alpha$ decay chains of their 4n and 3n evaporation residues, correspondingly.
The right panel shows the survival probability of the compound nuclei $^{283}$112 and $^{287}$114 formed in the fusion reactions $^{48}$Ca+$^{235}$U
(dashed curves) and $^{48}$Ca+$^{239}$Pu (solid curves).\label{Bf_Bn}}
\end{figure}

It is more convenient (and easier) to darn the gap ``from above'' by synthesis of new isotopes of SH elements
with larger values of Z, their subsequent $\alpha$ decay chains just fill the gap. This unexpected finding is simply explained by
greater values of survival probabilities of the corresponding nuclei with Z=115, 116 as compared to those with Z=111, 112.
In the left panel of Fig.\ \ref{Bf_Bn} the values of $B_f-B_n$ are shown for the SH mass area, where $B_f$ is the fission barrier and $B_n$
is the neutron separation energy (an odd-even effect is smoothed here). The values of $B_f-B_n$ are much higher
for CN with Z$\sim 116$ as compared with isotopes of element 112 formed in fusion reactions of $^{48}$Ca
with neutron deficient isotopes of uranium. As a result, the corresponding survival probability of lighter CN is smaller by more
than one order of magnitude. The right panel of Fig.\ \ref{Bf_Bn} shows survival probabilities of two compound nuclei,
$^{283}$112 and $^{287}$114, formed in the fusion reactions $^{48}$Ca+$^{235}$U and $^{48}$Ca+$^{239}$Pu.
The excitation energies of both compound nuclei (at collision energies equal to the corresponding Bass barriers, 195 and 198~MeV, correspondingly)
are just the same for two reactions (they are about 30~MeV).

In spite of the decrease of the fusion probability with increasing
charge number of the target nucleus, we may conclude that the EvR cross sections for the $^{48}$Ca+$^{239}$Pu reaction
should be higher (by about one order of magnitude for the 3n evaporation channel) due to the larger survival probability of $^{287}$114
compound nucleus as compared to $^{283}$112. Numerical calculations fully confirm this conclusion.
Thus the new isotopes of element 112 (at least, $^{279,280}$112) could be easier synthesized and studied as $\alpha$ decay products
of the heavier elements, 114 and/or 116.

\begin{figure}[ht]
\begin{center}
\includegraphics[width = 15.0 cm, bb=0 0 2402 594]{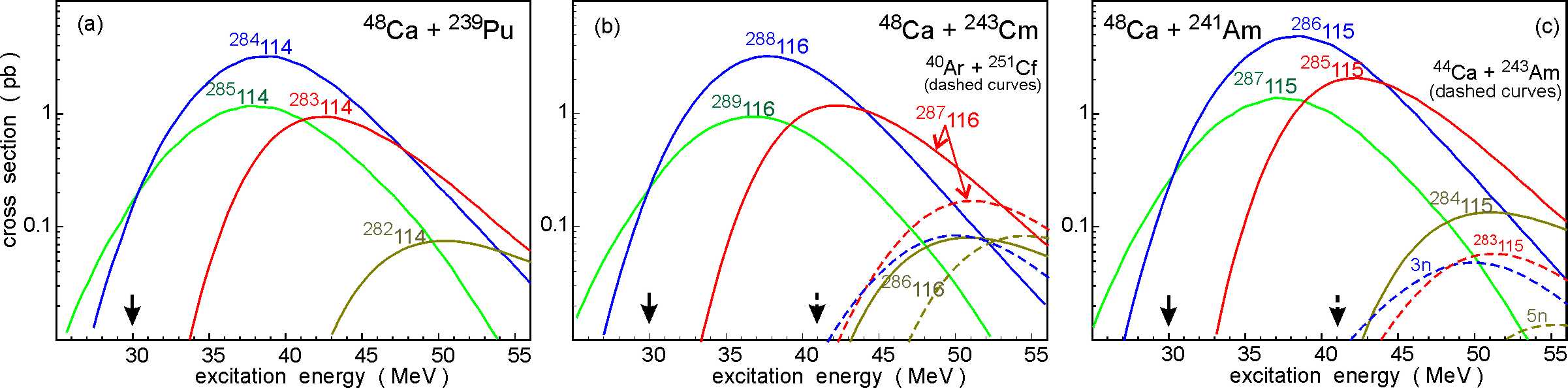} \end{center}
\caption{Production cross sections for the new isotopes of elements 114 (a), 116 (b) and 115 (c) in the $^{48}$Ca+$^{239}$Pu,
$^{48}$Ca+$^{243}$Cm, $^{40}$Ar+$^{251}$Cf (dashed curves), $^{48}$Ca+$^{241}$Am and
$^{44}$Ca+$^{243}$Am (3n, 4n and 5n evaporation channels, respectively, dashed curves) fusion reactions.
The arrows show positions of the corresponding Bass barriers.\label{114_116}}
\end{figure}

In Fig.\ \ref{114_116} the calculated EvR cross sections are shown for the production of new isotopes of elements 114 and 116
in the $^{48}$Ca+$^{239}$Pu, $^{48}$Ca+$^{243}$Cm and $^{40}$Ar+$^{251}$Cf fusion reactions.
High intensive beam of $^{40}$Ar can be obtained quite easily. This material is also much cheaper than $^{48}$Ca.
However, as can be seen from Fig.\ \ref{114_116}, the use of an $^{40}$Ar beam is less favorable as compared with $^{48}$Ca.
This is due to much ``hotter'' character of the $^{40}$Ar+$^{251}$Cf fusion reaction (only the cross sections for the 5n
evaporation channels are comparable for both reactions).
More than ten new isotopes of even elements from $Z=104$ to 116 could be produced in the $^{48}$Ca+$^{239}$Pu and/or $^{48}$Ca+$^{243}$Cm
fusion reactions which just fill the gap in the superheavy mass area. Note that the production cross sections are high enough to perform
such experiments at available facilities. All the decay chains, most probably, reach finally known nuclei (see Ref.\ \cite{Zag12}).
This fact significantly facilitates the identification of the new SH isotopes.

The $^{48}$Ca+$^{241}$Am fusion reaction is the best for the production of the new isotopes of odd SH elements filling the gap.
The production cross sections for the new isotopes $^{284-286}$115 in this reaction are about 0.1~pb, 2~pb and 4~pb, respectively,
i.e. high enough to be measured. The corresponding excitation functions are shown in Fig.\ \ref{114_116}.
The more neutron deficient isotopes of element 115 could be produced in the $^{44}$Ca+$^{243}$Am fusion reaction (note that $^{44}$Ca
is a more abundant and available material as compared to $^{48}$Ca). However in this reaction the excitation energy of the formed CN is
10~MeV higher than in the $^{48}$Ca+$^{241}$Am fusion reaction. As a result, the corresponding excitation functions (see the dashed curves
in Fig.\ \ref{114_116}(c)) are shifted to higher energies at which the survival probability of the CN is much lower.
Thus, the $^{48}$Ca beam remains preferable also for the production of neutron deficient SH nuclei in fusion reactions
with lighter isotopes of actinide targets as compared to the use of $^{42-44}$Ca or $^{40}$Ar beams.

Recently the synthesis of SH elements at the level of 1 pb became more or less a routine matter for several laboratories. The
corresponding experiments require about 2-week irradiation time to detect several decay chains of SH element.
This means that many new isotopes of SH elements could be synthesized now, and the gap between
nuclei produced in the cold and hot fusion reactions could be closed at last.
It can be done with the use of ordinary fusion reactions and, thus, with the use of existing recoil separators.

\subsection{Narrow pathway to the Island of Stability}

It is well known that there are no combinations of available projectiles and targets, the fusion of which may lead to SH
nuclei located at the island of stability. Only the proton-rich isotopes of SH elements have been produced so far in fusion
reactions (see Fig.\ \ref{mapup1}). Radioactive ion beams may hardly solve this problem. Fusion cross sections for relatively light
radioactive projectiles (such as $^{22}$O, for example) are rather high and a beam intensity of about $10^8$ pps is sufficient
for synthesis of SH elements \cite{Zag08}. However the nuclei, being synthesized in such a way, would be also neutron deficient.
For example, in the $^{22}$O + $^{248}$Cm fusion reaction one may produce only already known neutron-deficient isotopes of rutherfordium, $^{265-–267}$Rf.
In fusion reactions with heavier radioactive projectiles (such as $^{44}$S, for example) new neutron-enriched isotopes of
SH elements could be really produced, but in this case one needs to have a beam intensity of about $10^{12}$ pps to reach
in experiment a 1 pb level of the corresponding EvR cross section \cite{Zag08}, which is not realistic for the near future.

Still several more neutron rich actinide targets ($^{250}$Cm, $^{251}$Cf, $^{254}$Es) could be used, in principle, for production
of SH nuclei shifted by one or two neutrons to the right side from those already synthesized in $^{48}$Ca induced fusion reactions
(though they will be far from the beta-stability line, see Fig.\ \ref{mapup1}).
The EvR cross sections for the synthesis of elements 116, 118 and 119 formed in fusion reactions of $^{48}$Ca with
$^{250}$Cm, $^{251}$Cf and $^{254}$Es targets have been calculated in Ref.\ \cite{Zag12}.
As mentioned above, the $^{254}$Es target is rather exotic and hardly may be prepared, but a quite sufficient amount of the isotope
$^{251}$Cf (T$_{1/2}$ = 898 yr) is accumulated in nuclear reactors, and the only problem is its separation.

\begin{figure}[ht]
\begin{center}\includegraphics[width = 11.5 cm, bb=0 0 1972 1368]{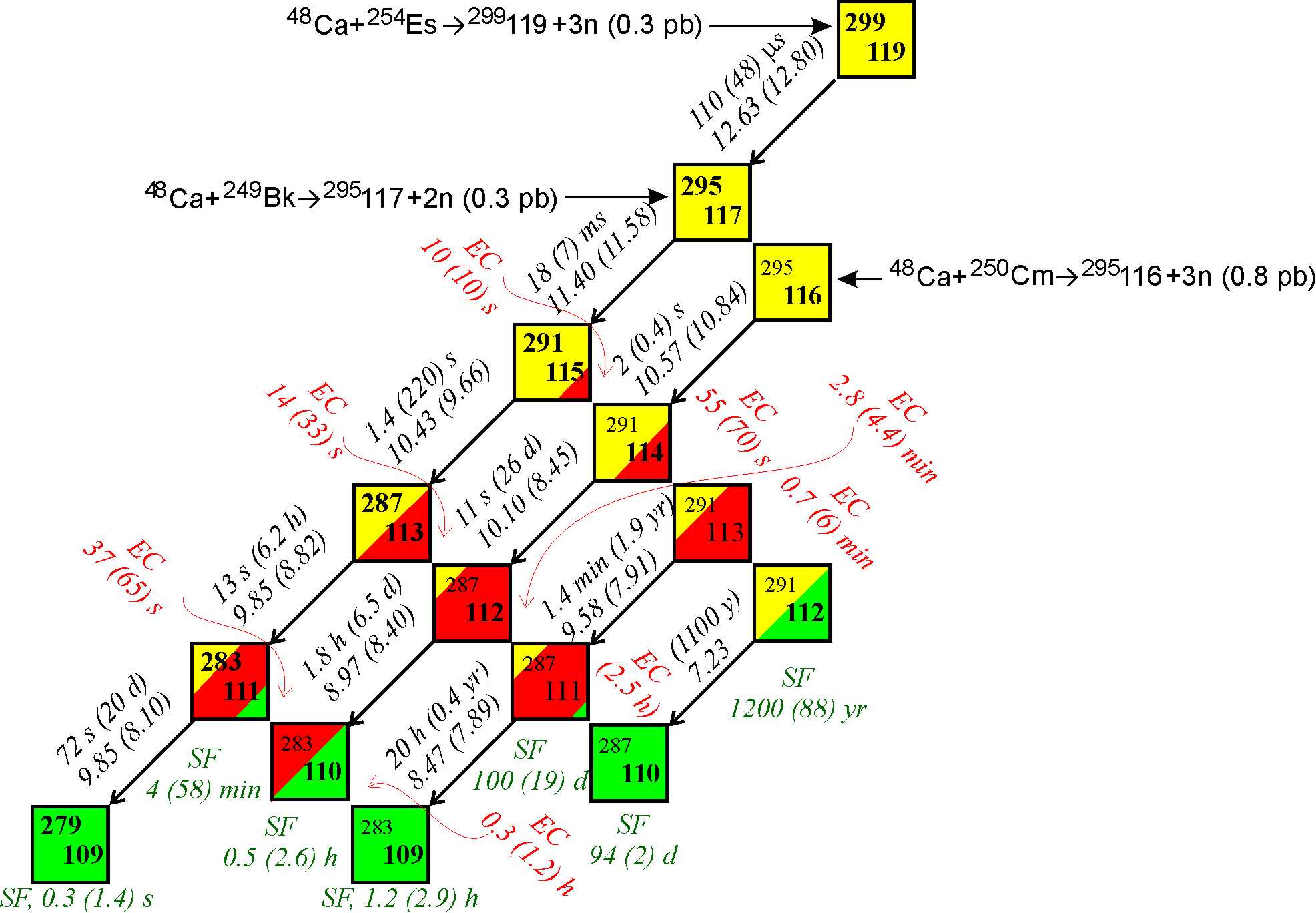}\end{center}
\caption{The pathway to the middle of the island of stability via a possible $\beta^+$ decay of the isotopes $^{291}$115 and  $^{291}$114.
The first isotope may be formed after $\alpha$ decay of $^{295}$117 (2n channel of the $^{48}$Ca+$^{249}$Bk fusion reaction,
cross section is 0.3~pb \cite{Zag08}) or after two  $\alpha$ decays of $^{299}$119 (3n, $^{48}$Ca+$^{254}$Es).
The second one, $^{291}$114, is formed after $\alpha$ decay of $^{295}$116 in the 3n evaporation channel of the $^{48}$Ca+$^{250}$Cm
fusion reaction with cross section of about 0.8~pb. \label{beta_decay}}
\end{figure}

New neutron rich isotopes of elements 116 ($^{294,295}$116) and 118 ($^{295,296}$118) may be synthesized in 3n and 4n
evaporation channels of the $^{48}$Ca+$^{250}$Cm and  $^{48}$Ca+$^{251}$Cf fusion reactions with the cross sections of about 1~pb.
Subsequent $\alpha$-decays of the nuclei $^{295,296}$118 pass through the known isotopes of elements 116, 114 and so on. It significantly
facilitates their detection and identification. $\alpha$ decay chains of $^{294}$116 and $^{295}$116 nuclei lead to absolutely new neutron
enriched isotopes of SH elements ended by fission of seaborgium and/or rutherfordium isotopes located already at the beta--stability line.
The cross section for production of element 119 in the $^{48}$Ca+$^{254}$Es fusion reaction is rather low ($\sim$0.3~pb)
but still it is larger than the cross section of the $^{50}$Ti+$^{249}$Cf fusion reaction which was estimated in \cite{Zag08} to be about 0.05~pb.

Another interesting feature of the fusion reactions $^{48}$Ca+$^{250}$Cm and $^{48}$Ca+$^{254}$Es (as well as the 2n evaporation channel
of the reaction $^{48}$Ca+$^{249}$Bk) is an unexpected possibility to reach the middle of the island of stability
just in fusion processes of ``stable'' nuclei. In these reactions relatively neutron rich isotopes of SH elements 114 and 115 are formed
as $\alpha$ decay products of evaporation residues of the corresponding CN. These isotopes should have rather long half-lives
and, thus, they could be located already in the ``red'' area of the nuclear map, i.e., they may be $\beta^+$-decaying nuclei.
In Fig.\ \ref{beta_decay} several possible decay chains of these isotopes are shown along with the corresponding values
of $Q_\alpha$ and half-lives calculated with the use of nuclear masses predicted by A.~Sobizcewski et al. \cite{Sob03}.
and by P.~M\"{o}ller et al. \cite{Moller97}. Spontaneous fission half-lives are taken from \cite{Smol97} (with the hindrance factor 100 for odd nuclei),
while the values in brackets are calculated by phenomenological relations \cite{Karpov12} with the shell corrections taken from \cite{Moller97}.

In accordance with our calculations of decay properties of SH nuclei \cite{Karpov12}, the isotopes $^{291}$115 and  $^{291}$114
may experience not only $\alpha$ decay but also electron capture with half-life of several seconds.
If it is correct, the narrow pathway to the middle of the island of stability is suprisingly opened by production of these isotopes
in subsequent $\alpha$-decays of elements 116, 117 and/or 119 produced in the $^{48}$Ca+$^{250}$Cm, $^{48}$Ca+$^{249}$Bk
and $^{48}$Ca+$^{254}$Es fusion reactions, see Fig.\ \ref{beta_decay}. The corresponding cross sections of these reactions are rather low,
they are about 0.8~pb for the 3n evaporation channel of the $^{48}$Ca+$^{250}$Cm fusion reaction and 0.3~pb for the two last reactions \cite{Zag12,Zag08}.
However, for the moment, this is the only method which is proposed for the production of SH nuclei located just in the middle of the island of stability.
Further careful study of the decay properties of unknown SH nuclei located closer to the beta-stability line is needed to confirm
the existence of such a possibility.

\subsection{Production of new neutron rich SH nuclei in transfer reactions}

The multi-nucleon transfer processes in near barrier collisions of heavy ions, in principle, allow one to produce
heavy neutron rich nuclei including those located at the island of stability. These reactions were studied extensively
about thirty years ago. Among other topics, there had been great interest in the use of heavy-ion transfer reactions
to produce new nuclear species in the transactinide region \cite{Hulet77,Rehm79,Freies79,Schadel82,Moody86,Welch87}.
The cross sections were found to decrease very rapidly with increasing atomic number of surviving heavy fragments.
However, Fm and Md neutron rich isotopes have been produced at the level of 0.1~$\mu$b \cite{Schadel82}.
It was observed also that nuclear structure (in particular, the closed neutron shell $N=126$) may influence nucleon flow
in dissipative collisions with heavy targets \cite{Mayer85}.

In our previous studies we found that the shell effects (clearly visible in fission and quasi-fission processes)
also play a noticeable role in near barrier multi-nucleon transfer reactions \cite{ZG08prl,ZG07b}.
These effects may significantly enhance the yield of searched-for neutron rich heavy nuclei for appropriate
projectile--target combinations. In particular, the predicted process of anti-symmetrising (``inverse'') quasi-fission
may significantly enhance the yields of long-living neutron rich SH isotopes in collisions of actinide nuclei.
However, the role of the shell effects in damped collisions of heavy nuclei is still not absolutely clear and was not carefully studied experimentally.
Very optimistic experimental results were obtained recently \cite{Loveland11} confirming such effects in the $^{160}$Gd+$^{186}$W reaction,
for which the similar ``inverse quasi-fission'' process ($^{160}$Gd$\to ^{138}$Ba while $^{186}$W$\to ^{208}$Pb) has been also predicted \cite{ZG07b}.

\begin{figure}[ht]
\begin{center}\includegraphics[width = 15.0 cm, bb=0 0 2265 558]{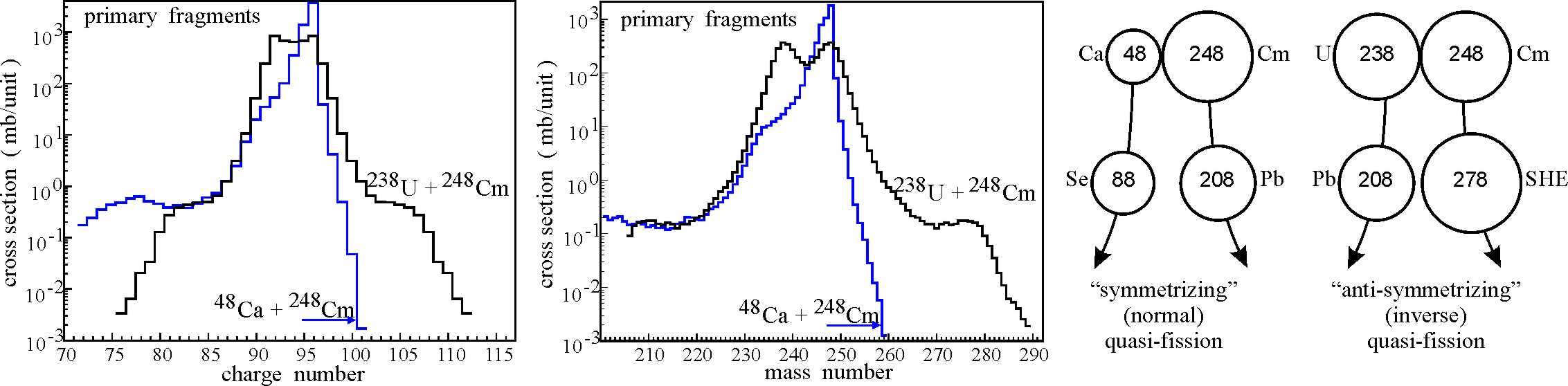}\end{center}
\caption{Charge and mass distributions of heavy primary reaction fragments formed in collisions of $^{238}$U and $^{48}$Ca with $^{248}$Cm target
at $E_{\rm c.m.}$=750 and 220 MeV, correspondingly. Schematic view of ``normal'' and ``inverse'' quasi-fission processes are also shown. \label{uca}}
\end{figure}

In multi-nucleon transfer reactions the yields of SH elements with masses heavier than masses of colliding nuclei strongly depend on the reaction combination.
The cross sections for the production of {\it neutron rich} transfermium isotopes in reactions with $^{248}$Cm target
change sharply if one changes from medium mass (even neutron rich) projectiles to the uranium beam. In Fig.\ \ref{uca} the charge and mass distributions
of heavy primary reaction fragments are shown for near barrier collisions of $^{48}$Ca and $^{238}$U with curium target.
The ``lead shoulder'' manifests itself in both reactions. However, for $^{48}$Ca+$^{248}$Cm collisions
it corresponds to the usual (symmetrizing) quasi-fission process in which nucleons are transferred mainly from the heavy target
(here it is $^{248}$Cm) to the lighter projectile. This is a well studied process both experimentally \cite{Itkis04} and
theoretically \cite{Zag05}. It is caused just by the shell effects leading to the deep lead valley on the multi-dimensional potential
energy surface which regulates the dynamics of the heavy nuclear system at low excitation energies (see Fig.\ \ref{adiabat}).

Contrary to this ordinary quasi-fission phenomena, for the $^{238}$U+$^{248}$Cm collisions nucleons are predominantly transferred from the lighter partner
(here is uranium) to heavy one (i.e. U transforms to Pb and Cm to 106 element). In this case, besides the lead shoulder in the mass and charge
distributions of the reaction fragments, there is also a pronounced shoulder in the region of SH nuclei (see Fig.\ \ref{uca}).

\begin{figure}[ht]
\begin{center}\includegraphics[width = 12.0 cm, bb=0 0 1992 997]{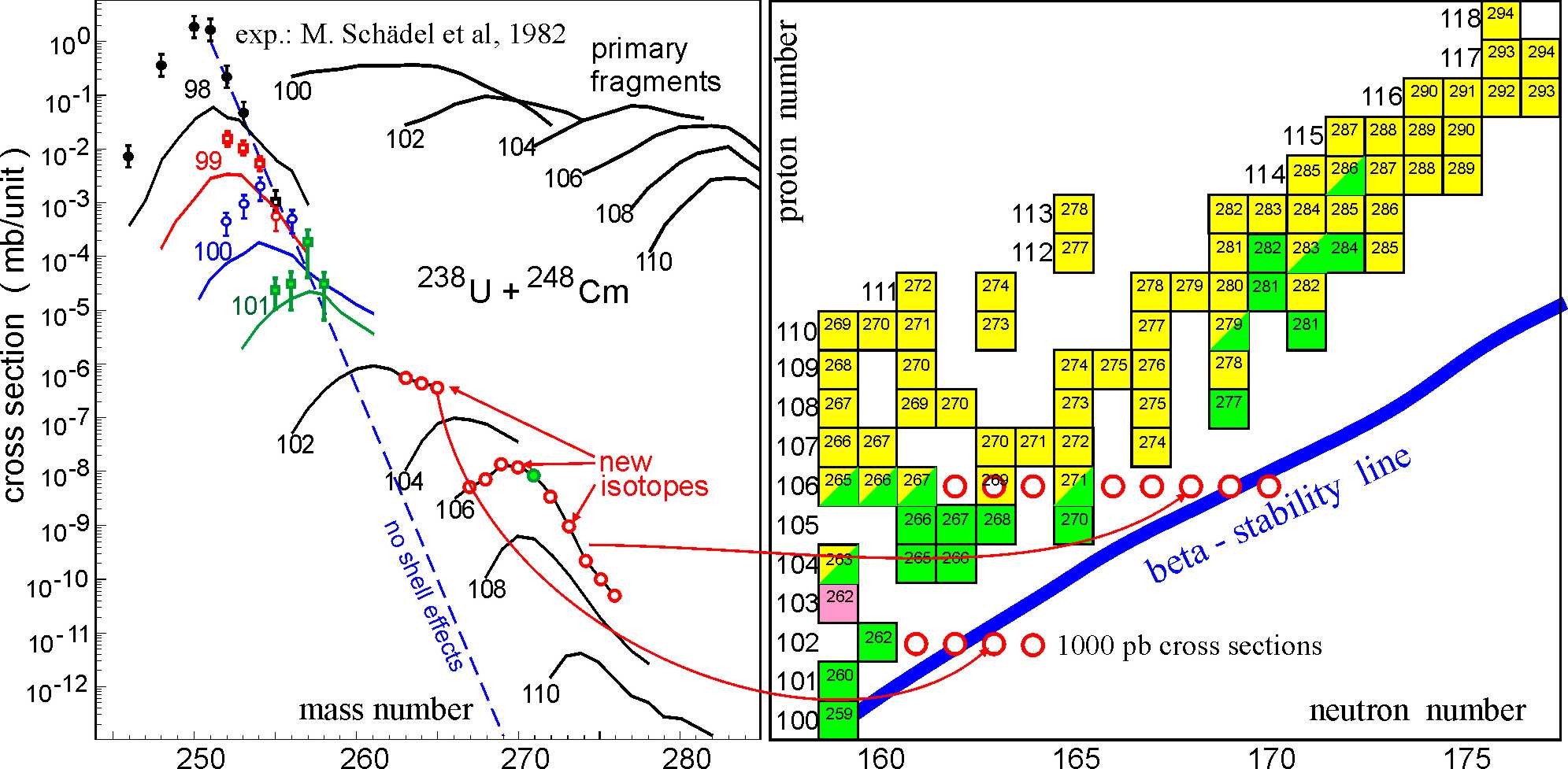}\end{center}
\caption{Yield of survived isotopes of SH nuclei produced in collisions of $^{238}$U with $^{248}$Cm target at $E_{\rm c.m.}$=750~MeV.
New isotopes of nobelium and siborgium are shown by open circles.\label{ucmcs}}
\end{figure}

Of course, the yield of survived SH elements produced in the low-energy collisions of actinide nuclei is rather low, though the shell effects give us
a definite gain as compared to a monotonous exponential decrease of the cross sections with increasing number of transferred nucleons.
In Fig.\ \ref{ucmcs} the calculated EvR cross sections for the production of SH nuclei in damped collisions of $^{238}$U with $^{248}$Cm
at 750~MeV center-of-mass energy are shown along with available experimental data. As can be seen, really many new neutron-rich isotopes
of SH nuclei with $Z>100$ might be produced in such reactions.

The choice of collision energy is very important for the production of desired neutron-rich SH nuclei.
With increasing beam energy the yield of primary fragments increases. However the excitation energy of these fragments
also increases and thus decreases their survival probabilities. We found that the optimal beam energy for the production of
neutron-rich isotopes of SH elements in multi-nucleon transfer reactions with heavy actinide nuclei (like U+Cm) is very
close to the energy needed for these nuclei to reach the contact configuration (there is no ordinary barrier: the potential energy
of these nuclei is everywhere repulsive). For $^{238}$U+$^{248}$Cm it is about 750~MeV center-of-mass collision energy.

\subsection{Synthesis of SH nuclei by neutron capture (SH elements in nature)}

\begin{figure}[ht]
\begin{center}\includegraphics[width = 12.5 cm, bb=0 0 2284 782]{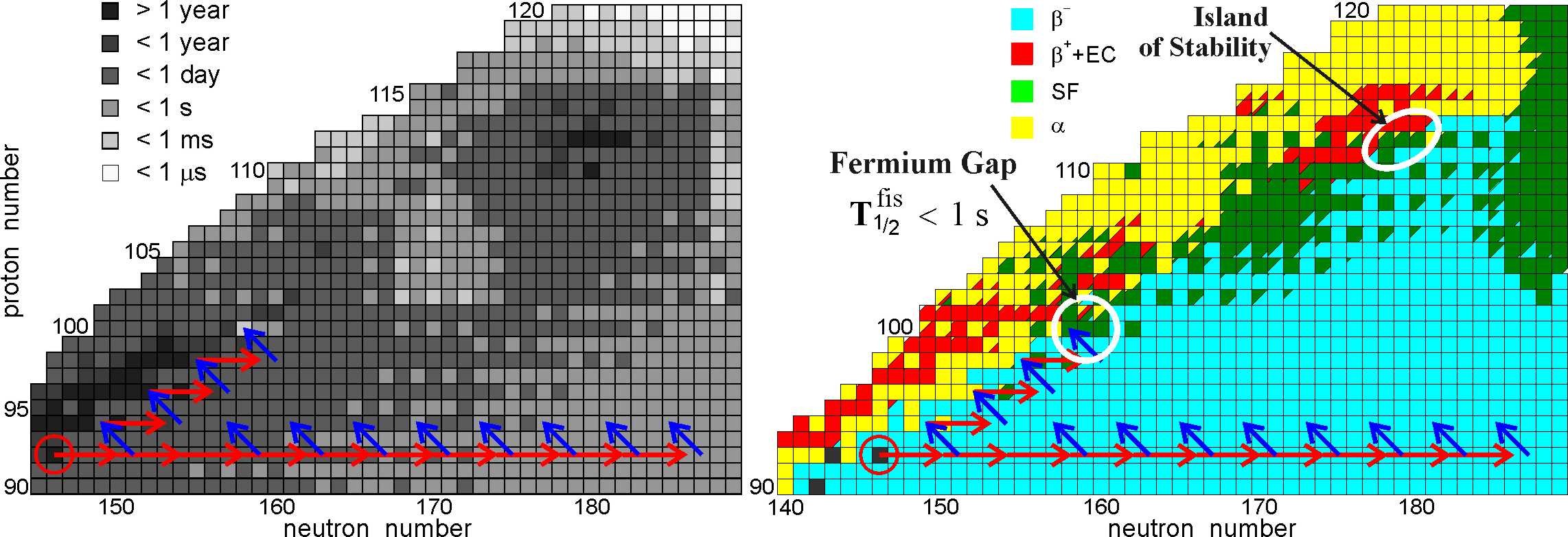}\end{center}
\caption{Calculated half-lives and preferable modes of decay of nuclei in the upper part of the nuclear map.
Schematic views of slow (terminated at the short-living fission fermium isotopes) and fast neutron capture processes with
subsequent $\beta^-$ decays are shown by arrows.\label{maphl}}
\end{figure}

The neutron capture process is an alternative (oldest and natural) method for the production of new heavy elements.
Strong neutron fluxes might be provided by nuclear reactors and nuclear explosions under laboratory conditions and by
supernova explosions in nature. The ``fermium gap,'' consisting of the short-living isotopes
$^{258-260}$Fm located on the $\beta$--stability line and having very short half-lives for spontaneous fission, impedes the formation of
nuclei with Z$>$100 by the weak neutron fluxes realized in existing nuclear reactors.
Theoretical models predict also another region of short-living nuclei located at Z=104$\div$108 and A$\sim$275 (see Fig.\ \ref{maphl}).

In nuclear and supernova explosions (fast neutron capture) these gaps may be bypassed if the total neutron fluence is high enough.
Note that elements 99 and 100 (einsteinium and fermium) were first discovered in debris of the test thermonuclear explosion ``Mike'' \cite{Mike}.
The experimental data on the yields of transuranium nuclei formed in this explosion are reproduced quite well with Eqs.(\ref{equation}) \cite{Zag11b}.

The resulting charge number of the synthesized nuclei might be increased by sequential neutron flux exposure if two or several nuclear explosions
were generated in close proximity to each other. This natural idea was discussed many years ago \cite{Meldner72}. At that time the experts (such as
Edward Teller) concluded that technically it could be realized. However, no quantitative estimations have been done for the yields of
SH neutron rich nuclei in such processes.

\begin{figure}[h]
\begin{center}\includegraphics[width = 12.0 cm, bb=0 0 1862 625]{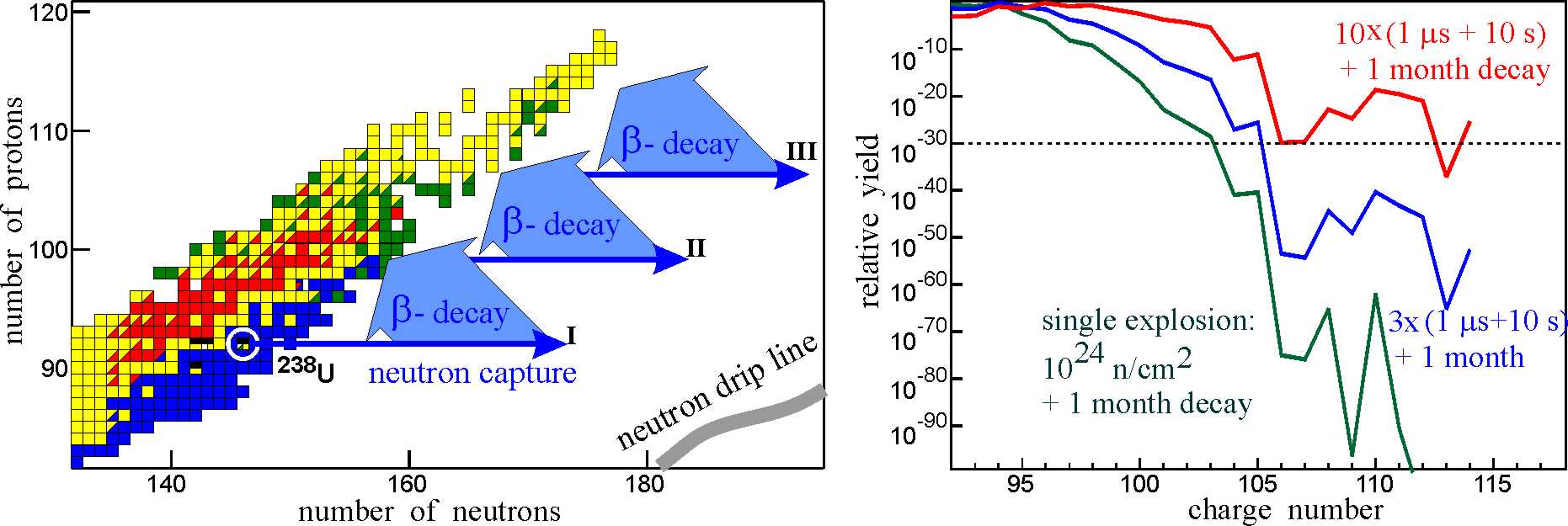}\end{center}
\caption{Schematic picture for multiple neutron irradiation of initial $^{238}$U material (left)
and probability for formation of heavy nuclei right) in such processes (one, three and ten subsequent explosions).
The dotted line denotes the level of few atoms.\label{multiple}}
\end{figure}

This process is illustrated in Fig.\ \ref{multiple}.
In the right panel of this figure the probabilities of heavy element formation are shown for one, three and ten subsequent
short-time ($1\,\mu$s) neutron exposures of $10^{24}$~n/cm$^2$ each following one after another within a time interval of 10 seconds
with final one-month waiting time (needed to reduce the strong radioactivity of the produced material and to perform some experimental measurements).
The result depends both on the neutron fluence $n=n_0\tau$ and on the time interval between two exposures.
The neutron fluence should be high enough to shift the produced neutron rich isotopes to the right from the second gap
of unstable fissile nuclei located at Z=104$\div$108 and A$\sim$275 (see Fig.\ \ref{maphl}).
Dependence on the time interval between two exposures is not so crucial.
It must be longer than several milliseconds (to avoid approaching
the neutron drip line after several exposures) and shorter than a few minutes to avoid $\beta^-$-decay of the produced
nuclei into the area of fission instability.

The same process of multiple neutron exposures might be also realized in pulsed nuclear reactors. Here the pulse duration is much longer
than in nuclear explosions (up to several milliseconds). However, the neutron fluence usually does not exceed $10^{16}$ n/cm$^2$ in existing nuclear reactors ($n_0\sim 10^{19}$~n/cm$^2$s during one millisecond pulse). Thus, the time of neutron capture $\tau_n = (n_0\sigma_{n\gamma})^{-1}\sim 10^5$~s,
and only the nearest long living isotopes (A+1 or A+2) of irradiated elements can be formed during the pulse. Multi--pulse irradiation here corresponds,
in fact, to the ``slow'' neutron capture process, in which new elements with larger charge numbers are situated close to the line of stability
and finally reach the fermium gap where the process stops. In this case the probability for formation of heavy elements
with $Z>100$ is negligibly small independent of the number of pulses and total time of irradiation.

The situation may change if one could be able to increase somehow the intensity of the pulsed reactor. The neutron fluence of one pulse and frequency
of pulses should be high enough to bypass both gaps of short living nuclei on the way to the island of stability (see Fig.\ \ref{maphl}).
The specifications of the high--intensity pulsed reactors of the next generation depends strongly on properties of heavy neutron rich nuclei
located to the right of these gaps. Using our theoretical estimations for the decay properties of these nuclei we have found that
increase of the neutron fluence in the individual pulse by about three orders of magnitude as compared with existing pulsed reactors,
i.e. up to 10$^{20}$ neutrons/cm$^2$, could be quite sufficient to bypass both gaps \cite{Zag11b}.

The astrophysical {\it r} process of nucleosynthesis is usually discussed to explain the observed abundance of heavy elements in the universe.
In such a process some amount of SH elements of the island of stability might be also produced if the fast neutron flux is sufficient to bypass
the two gaps of fission instability mentioned above. Strong neutron fluxes are expected to be generated by neutrino-driven proto-neutron star
winds which follow core-collapse supernova explosions \cite{Thompson2001} or by the mergers of neutron stars \cite{Rosswog99}.
Estimation of relative yields of SH elements is a difficult problem which depends both on the features of neutron fluxes and on the experimentally unknown
decay properties of heavy neutron rich nuclei. We mention only one of a very few such calculations made recently \cite{Panov09},
which gives the ratio of the yields of SH elements and uranium Y(SH)/Y(U)=$10^{-2}-10^{-20}$.

\begin{figure}[h]
\begin{center}\includegraphics[width = 13.0 cm, bb=0 0 2251 1065]{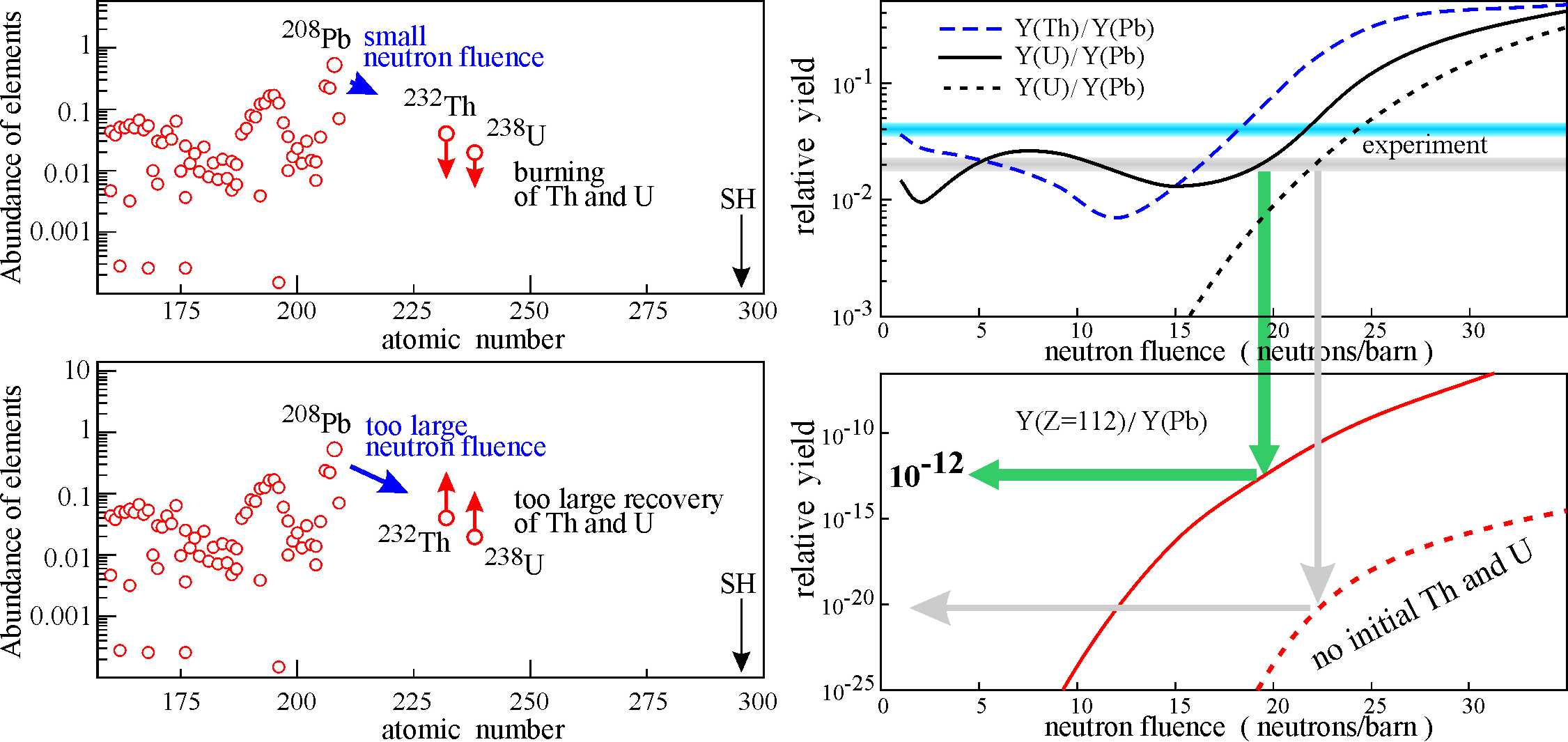}\end{center}
\caption{Initial relative abundance of nuclei (left panels). Burning and recovery of Th and U nuclei are shown schematically.
Relative to lead yields of thorium (dashed) and uranium (solid curve) nuclei depending on total neutron fluence in the astrophysical
{\it r} process are shown on the top right panel. The horizontal bars show experimental values of thorium and uranium abundances.
The same for relative yield of long-living SH copernicium isotopes $^{291}$Cn and $^{293}$Cn (right bottom panel).
The dotted curves show the yields of uranium and SH elements in the case of zero initial abundance of thorium
and uranium at the beginning of the {\it r} process.\label{sn}}
\end{figure}

We made a very simple estimation of the possibility for formation of SH nuclei during
the astrophysical {\it r} process of neutron capture. This estimation is based on the following assumptions.
(1) SH nuclei are relatively short-living. They are absent in stars initially, while the distribution of other elements
is rather close to their abundance in the universe.
(2) SH nuclei may appear at the last (rather cold) stage of the astrophysical {\it r} process
when the observed abundance of heavy elements (in particular, thorium and uranium to lead ratios) is also reproduced.
(3) Existing (experimental) abundance of stable nuclei may be used as initial condition. During intensive neutron irradiation
initial thorium and uranium material are depleted transforming to heavier elements and going to fission,
while more abundant lead and lighter stable elements enrich thorium and uranium.
(4) Unknown total neutron fluence may be adjusted in such a way that the ratios Y(Th)/Y(Pb) and Y(U)/Y(Pb)
keep its experimental values at the end of the process. Simultaneously, for a given neutron fluence,
one gets the relative yield of SH elements, Y(SH)/Y(Pb).

The results of our calculations are presented in the right panels of Fig.\ \ref{sn}, where the relative to lead yields of thorium, uranium and
long-living SH copernicium isotopes are shown depending on the total neutron fluence.
At low neutron fluxes initial thorium and uranium nuclei increase their masses and charges
(after neutron capture and subsequent $\beta^-$-decay), find themselves in the region of fission instability and drop out.
Thus, their numbers decrease relative to lead, which, in contrast with Th and U, has an additional feed from lighter nuclei.
At neutron fluence  $n \sim 1.5\cdot 10^{25}$~cm$^{-2}$ (= 15 neutrons/barn) burning of thorium and uranium
is compensated by increasing contribution from lighter stable nuclei with $Z\le 83$,
and at $n \sim 2\cdot 10^{25}$~cm$^{-2}$ are both ratios, Y(Th)/Y(Pb) and Y(U)/Y(Pb), close to the observed values.
At this neutron fluence the relative to lead yield of most stable isotopes of SH element 112,
namely $^{291}$Cn and $^{293}$Cn, is about $10^{-12}$ which is not extremely low and keeps hope to find them in nature (most probably in the cosmic rays).
If one assumes that initial thorium and uranium nuclei are completely burned in s-process of neutron capture before supernova explosion
then the yield of SH nuclei is by about 8 orders of magnitude less than in the former case (see dashed curves in Fig.\ \ref{sn}).

\section{Summary}

First, we hope that new SH elements 119 and 120 will be successfully synthesized within one or two nearest years.
Synthesis of SH elements with Z$>$120 is rather problematic in near future due to extremely low cross sections and short half-lives of these elements.
One might think that the epoch of $^{48}$Ca in the production of SH nuclei was finished by the synthesis of element 118 in the $^{48}$Ca+$^{249}$Cf
fusion reaction \cite{118}. However this projectile still could be successfully used for the production of new isotopes of SH elements.
The extension of the area of known isotopes of SH elements is extremely important for better understanding of their properties
and for developing the models which will be able to predict well the properties of SH nuclei located beyond this area
(including those at the island of stability). The ordinary fusion reactions could be used for the production
of new isotopes of SH elements. The gap of unknown SH nuclei, located between the isotopes which were produced earlier in the ``cold''
and ``hot'' fusion reactions, could be filled in fusion reactions of $^{48}$Ca with available lighter isotopes of Pu, Am and Cm.

Then we must redirect our interests onto the production of longer living neutron enriched SH nuclei.
The low-energy multi-nucleon transfer reactions can be really used for the production and for the study of the properties of new
neutron rich isotopes of heavy elements in the upper part of the nuclear map (from $Z\sim 70$ and up to SH elements).
For the SH mass region the multi-nucleon transfer process remains the only reaction mechanism which allows one
to produce more neutron rich and longer living SH nuclei. Here neutron enriched isotopes of all the elements with $Z\ge 100$
are of great interest, because all known isotopes of these elements are located at the proton rich side of the beta-stability
line. The use of the heaviest target and projectile combinations gives a gain in the cross sections for the production
of the most neutron rich isotopes with masses greater than the masses of both colliding nuclei.

The neutron-enriched isotopes of SH elements may be also produced with the use of $^{48}$Ca beam if a $^{250}$Cm target would be prepared.
In this case we get a real chance to reach the island of stability due to a possible $\beta^+$ decay of $^{291}$114 and $^{287}$112
nuclei formed in this reaction.
The same path to the island of stability is opened also in the 2n evaporation channel of the $^{48}$Ca+$^{249}$Bk fusion reaction
leading to the isotope $^{291}$115 having a chance for $\beta^+$ decay.

A macroscopic amount of the long-living superheavy nuclei located at the island of stability may be produced in multiple (rather ``soft'') nuclear explosions.
This goal could be also reached by using the pulsed nuclear reactors
of the next generation, if the neutron fluence per pulse will be increased by about three orders of magnitude.
Our estimation of the possibility for the production of SH elements in the astrophysical {\it r} process
(namely, neutron rich copernicium isotopes $^{291}$Cn and $^{293}$Cn with half-lives longer than several tens of years)
is not completely pessimistic: their relative to lead yield could be about $10^{-12}$
if one assumes initial natural abundance of all the elements (including thorium and uranium) at the beginning
of the astrophysical r-process. This ratio is not beyond the experimental sensitivity for a search for SH elements in nature
(for example, in cosmic rays).


\section*{References}
\begin{thebibliography}{99}
\bibitem{Hofmann00} S.~Hofmann and G.~M\"{u}nzenberg, Rev. Mod. Phys. {\bf 72}, 733 (2000).

\bibitem{Morita07} K.~Morita, K.~Morimoto et al., J. Phys. Soc. Jpn. {\bf 76}, No. 4, 043201 (2007); ibid. {\bf 76}, No. 4, 045001 (2007).

\bibitem{Ogan07} Yu Oganessian, J. Phys. G {\bf 34} R165 (2007).

\bibitem{118} Yu.Ts.~Oganessian, V.K.~Utyonkov, Yu.V.~Lobanov et al., Phys. Rev. C{\bf 74}, 044602 (2006).

\bibitem{Z03} V.I.~Zagrebaev, M.G.~Itkis, and Yu.Ts.~Oganessian, Phys. At. Nucl. {\bf 66}, 1033 (2003).

\bibitem{Z04} V.I.~Zagrebaev, Nucl. Phys. A{\bf 734}, 164 (2004).

\bibitem{Moller97} P.~M\"oller, J.~R.~Nix, and K.-L.~Kratz, At. Data Nucl. Data Tables {\bf 66}, 131 (1997).

\bibitem{Sob03} I.~Muntian, Z.~Patyk, and A.~Sobiczewski, Phys. At. Nucl., {\bf 66}, 1015 (2003) [Yad. Fiz., {\bf 66}, 1051 (2003)].

\bibitem{114_5n} P.A. Ellison, K.E. Gregorich, J.S. Berryman et al., Phys. Rev. Lett. {\bf 105}, 182701 (2010).

\bibitem{GSI_114} J.M.~Gates, Ch.E. D\"{u}llmann, M.~Sch\"{a}del et al., Phys. Rev. {\bf C 83}, 054618 (2011).

\bibitem{GSI_116} S.~Hofmann, S.~Heinz, R.~Mann et al., GSI Scientific Report, 197 (2010).

\bibitem{Zag12} V.I.~Zagrebaev, A.V.~Karpov, and W.~Greiner, Phys. Rev. C {\bf 85}, 014608 (2012).

\bibitem{Zag08} V.I.~Zagrebaev and W.~Greiner, Phys. Rev. C {\bf 78}, 034610 (2008).

\bibitem{Zag11} V.I.~Zagrebaev and W.~Greiner, Phys. Rev. C {\bf 83}, 044618 (2011).

\bibitem{Zag11b} V.I.~Zagrebaev, A.V.~Karpov, I.N.~Mishustin and W.~Greiner, Phys. Rev. C {\bf 84}, 044617 (2011).

\bibitem{Zag02} V.I.~Zagrebaev et al., Phys. Rev. C {\bf 65}, 014607 (2002); Empirical CC Code of NRV, http://nrv.jinr.ru/nrv.

\bibitem{NRV} V.I.~Zagrebaev et al., Statistical Model Code of NRV, http://nrv.jinr.ru/nrv.

\bibitem{Toke85} J.~T\={o}ke, R.~Bock, G.X.~Dai et al., Nucl. Phys. A {\bf 440}, 327 (1985).

\bibitem{ZSG07} V.~Zagrebaev, V.V.~Samarin, and W.~Greiner, Phys. Rev. C {\bf 75}, 035809 (2007).

\bibitem{Umar08} A.S.~Umar, V.E.~Oberacker, and J.A.~Maruhn, Eur. Phys. J. A {\bf 37}, 245 (2008).

\bibitem{Simenel12} C.~Simenel, talk at NNC conference, San Antonio, Texas (2012).

\bibitem{extTCSM} V.~Zagrebaev et al., Phys. Part. Nucl., {\bf 38}, No. 4, 469 (2007).

\bibitem{Zag05} V.~Zagrebaev and W.~Greiner, J. Phys. G {\bf 31}, 825 (2005); J. Phys. G {\bf 34}, 1 (2007).

\bibitem{Dorn62} D.W.~Dorn, Phys. Rev. {\bf 126}, 693 (1962).

\bibitem{Seaborg} G.T.~Seaborg, Ann. Rev. Nucl. Sci. {\bf 18}, 53 (1968).

\bibitem{Karpov12} A.V.~Karpov, V.I.~Zagrebaev, Y.~Martinez Palenzuela et al., Int. J. Mod. Phys. E {\bf 21}, 1250013 (2012).

\bibitem{SHIP120} S.~Hofmann et al., private communication.

\bibitem{TASCA120} Ch.~D\"{u}llmann et al., private communication.

\bibitem{Smol97} R.~Smola\'{n}czuk, Phys. Rev. C {\bf 56}, 812 (2002).

\bibitem{Hulet77} E.K.~Hulet, R.W.~Lougheed, J.F.~Wild et al., Phys. Rev. Lett. {\bf 39}, 385 (1977).

\bibitem{Rehm79} H.~Essel, K.~Hartel, W.~Henning, P.~Kienle et al., Z. Physik A {\bf 289}, 265 (1979).

\bibitem{Freies79} H.~Freiesleben, K.D.~Hildenbrand, F.~P\"{u}hlhofer et al., Z. Phys. A {\bf 292}, 171 (1979).

\bibitem{Schadel82} M.~Sch\"{a}del, W.~Br\"{u}chle, H.~G\"{a}ggeler et al., Phys. Rev. Lett. {\bf 48}, 852 (1982).

\bibitem{Moody86} K.J.~Moody, D.~Lee, R.B.~Welch, K.E.~Gregorich, G.T.~Seaborg et al., Phys. Rev. C {\bf 33}, 1315 (1986).

\bibitem{Welch87} R.B.~Welch, K.J.~Moody, K.E.~Gregorich, D.~Lee, G.T.~Seaborg et al., Phys. Rev. C {\bf 35}, 204 (1987).

\bibitem{Mayer85} W.~Mayer, G.~Beier, J.~Friese, W.~Henning, P.~Kienle et al., Phys. Lett. B {\bf 152}, 162 (1985).

\bibitem{ZG08prl} V.~Zagrebaev and W.~Greiner, Phys. Rev. Lett. {\bf 101}, 122701 (2008).

\bibitem{ZG07b} V.~Zagrebaev and W.~Greiner, J. Phys. G {\bf 34}, 2265 (2007).

\bibitem{Loveland11} W.~Loveland, A.M.~Vinodkumar, D.~Peterson and J.P.~Greene, Phys. Rev. C {\bf 83}, 044610 (2011).

\bibitem{Itkis04} M.G.~Itkis, J.~\"{A}yst\"{o}, S.~Beghini et al., Nucl. Phys. {\bf A734}, 136 (2004).

\bibitem{Mike} H.~Diamond, P.R.~Fields, C.S.~Stevens et al., Phys. Rev. {\bf 119}, 2000 (1960).

\bibitem{Meldner72} H.W.~Meldner, Phys. Rev. Lett. {\bf 28}, 975 (1972).

\bibitem{Thompson2001} T.A.~Thompson, A.~Burrows and B.S.~Meyer, Astrophys. J. {\bf 562}, 887 (2001).

\bibitem{Rosswog99} S.~Rosswog, M.~Liebend\"{o}rfer, F.-K.~Thielemann et al., Astron. Astrophys. {\bf 341}, 499 (1999).

\bibitem{Panov09} I.V.~Panov, I.Yu.~Korneev and F.-K.~Thielemann, Phys. At. Nucl. {\bf 72}, 1026 (2009).
\end{thebibliography}
\end{document}